\documentstyle[12pt]{article}
\def\tr{\,{\rm tr}\,}
\def\pr{\psi_{r}}
\def\prt{\psi_{r}^{T}}
\def\phrt{\phi_{r}^{T}}
\def\prbt{\bar{\psi}_{r}^{T}}

\def\prb{\bar{\psi}_{r}}
\def\aor{\alpha_{1}^{\rm ort}}
\def\am{A_{\mu}}
\def\amt{A_{\mu}^{T}}
\def\an{A_{\nu}}
\def\ant{A_{\nu}^{T}}
\def\al{A_{\lambda}}
\def\as{A_{\sigma}}
\def\alt{A_{\lambda}^{T}}
\def\ast{A_{\sigma}^{T}}
\def\ls{i\bar{\psi}\gamma^\mu(\partial_\mu +\am)\psi}
\def\phr{\phi_{r}}
\def\la{\lambda^{a}}
\def\lb{\lambda^{b}}

\def\phrb{\bar{\phi}_{r}}
\def\be{\begin{equation}}
\def\ee{\end{equation}}
\def\bea{\begin{eqnarray}}
\def\eea{\end{eqnarray}}
\title{{\bf \mbox{} \\ On the SO(N) symmetry of the chiral SU(N) Yang--Mills
model}}
\vspace{.7cm} \author{ \mbox{} \\ S.A.Frolov
\thanks{Permanent address:\ Steklov Mathematical Institute, Moscow} \mbox{} \\
\vspace{0.4cm}
Laboratoire de Physique Th\'eorique et Hautes Energies
\thanks{\it Unit\'e associ\'ee au C.N.R.S. URA 280},
Paris
\thanks{L.P.T.H.E. Tour 16- 1er \'etage Universit\'e Pierre et Marie Curie
4, place Jussieu 75252 PARIS CEDEX 05 - FRANCE} \mbox{} \\
\vspace{.6cm}
A.A.Slavnov and C.Sochichiu
\vspace{-0.5cm}
\mbox{} \\
Steklov Mathematical Institute
\vspace{-0.1cm}
\mbox{}  \\
Vavilov st.42, GSP-1, 117966 Moscow, RUSSIA \date{}}

\begin{document}

\maketitle
\vspace{4.5cm}
\begin{abstract}

The posibility of quantizing the anomalous $SU(N)$ Yang--Mills model
preserving the symmetry under the orthogonal subgroup is indicated.  The
corresponding Wess--Zumino action (1-cocycle) possesses the additional
$SO(N)$ symmetry and can be expressed in terms of chiral fields taking values
in the homogeneous space $SU(N)/SO(N)$. The modified anomaly and the
constraints commutator (2-cocycle)
are calculated.
\end{abstract}
\vspace{2.0cm}
PAR-LPTHE 92-48 \hspace{8.0cm} DECEMBER 1992
\newpage
\section {Introduction}

It is known that chiral gauge models suffer from anomalies \cite{Ad,Bar,GJ}.
The calculation of  the anomalies was performed both in the framework of
perturbation
 theory  and by algebraic and geometric methods \cite{St,Zu,Fa,FS84}. From
 algebraic point of view the anomaly corresponds to infinitesimal 1-cocycle
 on a group $G$. The global 1-cocycle as was indicated by Faddeev and
Shatashvili \cite{Fa,FS84} is just the Wess--Zumino action
 \cite{WZ}, depending  on the chiral fields with values  in the group $G$.
The anomaly leads also to the appearence of an additional term in the
constraints commutator which is the infinitesimal 2-cocycle on the group
$G$ \cite{Fa,FS84,FS86}. It was argued that it may change the physical content
of the theory.

The particular form of anomaly, the corresponding Wess--Zumino term and
2-cocycle depend on the regularization used. Although the difference is a
local term it  may lead  to important physical consequences. In the two
dimensional case it was shown by Jackiw and  Rajaraman \cite{JR} that
different counterterms result in the different spectrum  of  the  model.
In  the anomalous case there  is no  regularization preserving the full gauge
symmetry of the  theory and therefore a priori there is no unique choice of
the particular form of anomaly. However  it seems natural to choose a form
keeping as much of  classical gauge symmetries  as possible. It is known  that
there exist some chiral  gauge groups and the fermion representations for
which the anomalies compensate, for example the orthogonal  groups
$SO(N)$, $N\neq 6$. Presumably for such groups  and  representations one can
construct an invariant regularization explicitly preserving the gauge
symmetry. In particular for  the spinor representations of the $SO(N)$ groups
such a regularization was proposed in ref.\cite{FSl}.

In the general case as we  have already mentioned it is impossible to
preserve the full gauge symmetry but one can construct a regularization
invariant with respect to some nonanomalous subgroup. In this paper we shall
consider the chiral  $SU(N)$ gauge model with  fermions in the fundamental
representation. In this case one can maintain  the invariance  under the
orthogonal subgroup $SO(N)$. The resulting anomaly and the Wess--Zumino action
differ  from the "standard" ones discussed in the papers mentioned above  by
local terms. The Wess--Zumino action possesses the  additional $SO(N)$
invariance leading to  vanishing of the  anomaly on the $SO(N)$ subgroup. That
means the Wess--Zumino action depends in  fact on the  chiral fields taking
values not  in the group $SU(N)$ as in the usual case, but in the homogeneous
space  $SU(N)/SO(N)$. Therefore the number of degrees of freedom is reduced.
Naturally the 2-cocycle is also changed and the constraints commutator is
anomaly free on the  $SO(N)$ subgroup.

The Wess--Zumino action over homogeneous spaces was considered in several
publications (see for example \cite{Wi,Wu}). However the authors of these
papers were interested in the effective Nambu--Goldstone actions describing
low energy QCD. In our case the Wess--Zumino action over homogeneous spaces
arises in the  process of quantization of chiral Yang--Mills model preserving
a nonanomalous subgroup. We hope that it may be of interest for analyzing the
spectrum of anomalous models.

In the second section we give two expressions for the Wess--Zumino action. The
first one depends on the chiral fields with values in the group $SU(N)$  and
has  the additional $SO(N)$ invariance. Then we introduce chiral fields on the
homogeneous space and rewrite the Wess--Zumino action in terms of these
fields. Using this action we calculate anomaly. In the third section we
apply the method proposed  in ref.\cite{AMF} to get the expression for the
2-cocycle, appearing  in the anomalous constraints commutator.

\section{The $SO(N)$ invariant Wess--Zumino action}

We consider the model described by the Lagrangian
\be
{\cal L}=\ls
\label{1}
\ee
where $\am$ is a $SU(N)$ Yang--Mills  field  which in this section  will be
considered as an  external one, $\psi \equiv \frac{1}{2}(1+\gamma_{5})\psi$
is a chiral  fermion in the fundamental representation.  The fundamental
representation of the Lie algebra $su(N)$ is  generated by the antihermitian
matrices $\la$ :
\be
\tr{\la\,\lb}=-\frac{1}{2}\delta_{ab};\quad   [\la,\lb]=f^{abc}\,
\lambda^{c} \label{2}
\ee
and  $\am=\am^{a}\,\la$ .

The gauge transformation looks as follows:
\bea
&&\am\,\rightarrow  \,\am^{g}={g}^{-1}\am
{g}+{g}^{-1}\partial_{\mu}{g}\nonumber\\
&&\psi\,\rightarrow \,\psi^{g}={g}^{-1}\,\psi,\qquad {g}\in {SU(N)}
\label{3}
\eea
There is no $SU(N)$ invariant regularization  for this Lagrangian however one
can write the $SO(N)$ invariant regularized Lagrangian of  the form:
\bea
&&{\cal L}=\ls+\sum_{r=1}^{2K-1}[{i}\prb \gamma ^\mu (\partial _\mu +\am
)\pr -M_{r}\prt C \pr -M_{r}\prb C \prbt] \nonumber \\
&&\quad+{i}\sum_{r=1}^{2K}[(-1)^r \phrb \gamma ^\mu (\partial  _\mu + \am
)\phr-\sum_{s=1}^{2K}(M_{rs}\phrt C\phi_s - M_{rs}\phrb C\bar{\phi}_s^T )]
\label{4}
\eea
Here $\pr $ are the anticommuting Pauli--Villars  spinors  and
$\phr $ are the commuting ones. $M_{rs}$ is an antisymmetric matrix. The
standard Pauli--Villars conditions are assumed. The matrix $C$ is the charge
conjugation matrix. The only  terms, which  are not invariant under the
gauge transformation (\ref{3}) of all fields, are the mass term for the
Pauli--Villars fields.  The mass term transforms as follows
\be
M_{r}\prb C\prbt \rightarrow \,M_{r}\prb Cgg^{T} \prbt .
\label{5}
\ee
One sees that for $g\in SO(N)$, $gg^{T}=1$ this mass term is invariant, and
therefore the regularization preserves the $SO(N)$ gauge invariance.

The Wess--Zumino action  $\aor (A,g)$ is defined  by usual  formula
\be
\hbox{e}^{{i}\aor (A,g)}=\frac{\det (\gamma^\mu(\partial_\mu
+\am^g))}{\det (\gamma^\mu(\partial_\mu +\am))} .
\label{6}
\ee
Where it is understood that the  determinant is calculated with the help of
regularization (\ref{4}) and necessary  counterterms are introduced. It
follows directly from  eq.(\ref{6}) that the Wess--Zumino action is a
1-cocycle and satisfies the condition:
\be
\aor (A,g_1)+\aor (A^{g_1},g_2)=\aor (A,g_1 g_2)\qquad (mod\, 2\pi).
\label{7}
\ee
Due to $SO(N)$ gauge invariance of the regularized Lagrangian (\ref{4}) the
Wess--Zumino action has the additional invariance
\be
\aor (A,gh)=\aor (A,g),
\label{8}
\ee
where $h\in SO(N)$.

Let us stress that in eq.(\ref{8}) the field $A$ is not transformed.
Eq.(\ref{8}) is a direct consequence of the invariance of the gauge
transformed mass term (\ref{5}) under the transformation \be g\rightarrow
gh,\qquad h\in SO(N).  \label{9} \ee Eq.(\ref{8}) expresses the hidden
symmetry of the Wess--Zumino action in our case. Hidden symmetries of this
type in connection with models on homogeneous spaces were  discussed in
 refs.\cite{CJ,EF,Wu}. It follows from eqs.(\ref{7},\ref{8}) that the
Wess--Zumino action vanishes if the chiral field $g$ belongs to the
orthogonal subgroup $SO(N)$
\be
\aor (A,h)=0,\qquad h\in SO(N)
\label{10}
\ee
The geometric origin of the existence of such Wess--Zumino action is the
triviality of the cohomology group $H^5(SO(N))$.

The Wess--Zumino action depends on the regularization scheme used. The
difference is a trivial local 1-cocycle. We can use this fact to calculate
the Wess--Zumino action corresponding to regularization (\ref{4}) starting
from the action given for example in the paper \cite{FS84}. The action we are
interested in may be presented  in the form:
\be
\aor (A,g)=\alpha_1 (A,g) + \alpha_0 (A^g) -\alpha_0 (A).
\label{11}
\ee
Here $\alpha_1 (A,g)$ is the "standard" Wess--Zumino action
\bea
\alpha_1 (A,g)&=&\int d^4x\,[d^{-1}\kappa (g)-\frac {i}{48\pi^2}
\epsilon^{\mu \nu \lambda \sigma }\tr[(\am \partial_\nu \al
\,+\,\partial_\mu \an \al \,+\,\am \an \al )g_\sigma \,-\nonumber \\
&&-\,\frac 12\am g_\nu \al g_\sigma \,-\,\am g_\nu g_\lambda
g_\sigma]]
\label{12}
\eea
and we use the notations
\be
\int d^4xd^{-1}\kappa (g) \equiv -\frac{i}{240\pi^2}\int_{M_5} d^5x
\,\epsilon^{pqrst} \tr{(g_p
g_q g_r g_s g_t)}
\label{13}
\ee
\be
g_\mu=\partial_\mu gg^{-1}.
\label{14}
\ee
In eq.(\ref{13}) the  integration goes  over a  five-dimensional manifold
whose boundary is the usual  four-dimensional space.

The functional $\alpha_0 (A^g)- \alpha_0 (A)$ is a trivial local  1-cocycle
which can be determined  from eq.(\ref{10}). The explicit  form of $\alpha_1
(A,g)$ (eq.(\ref{12})) dictates the following  ansatz for  $\alpha_0 (A)$:
\bea
\alpha_0 (A)&=&-\frac {i}{48\pi^2}\int  d^4x\, \epsilon^{\mu \nu \lambda
\sigma } \tr(a_1 \am \an \al \ast+a_2 \am  \ant  \al \ast +\nonumber\\
&&+a_3 \am \an \alt \ast + b_1 \partial_\mu \an
\al \ast + b_2 \partial_\mu \an \alt \as + b_3 \partial_\mu \an \alt
\ast )
\label{15}
\eea
where $\amt$ is a transposed  matrix $\am $.

Let us stress that to satisfy eq.(\ref{10}) it is necessary to introduce the
terms depending not only on $\am$ but also on $\amt$. Eq.(\ref{10}) determines
uniquely the coefficients $a_i,b_i$. As a result:
\bea
\alpha_0 (A)&=&-\frac {i}{48\pi^2}\int  d^4x\, \epsilon^{\mu \nu \lambda
\sigma } \tr( \am \an \al \ast-\frac 14 \am  \ant  \al \ast +\nonumber\\
&& + \partial_\mu \an
\al \ast + \am \partial_\nu \al \ast )
\label{16}
\eea
Obviously one can add  also any trivial local $SO(N)$ invariant 1-cocycle.
The corresponding infinitesimal 1-cocycle (anomaly) is calculated in a
standard way
\be
\int d^4x\,\epsilon^a(x)\, {\cal A}^{a}_{ort}(A)=\aor (A^h,h^{-1}g)-\aor(A,g)
\label{17}
\ee
where $h=1+\epsilon^a \lambda^a$.

It looks as follows
\bea
&&{\cal A}^{a}_{ort}(A)=\frac {i}{48\pi^2}\epsilon^{\mu \nu \lambda \sigma }
\tr[( \lambda^a+\lambda^{a,T})(\partial_\mu (\an \partial_\lambda
\as+\nonumber \\
&&\quad+\partial_\nu \al \as+\an \al \as -\an \alt \as -\frac 12  \ant
\partial_\lambda\as-\frac 12\partial_\nu  \al \ast)-\nonumber\\
&&\quad-\partial_\mu \an \al \ast-\am \partial_\nu \al \ast-\amt \partial_\nu
\al
\as-\amt \an\partial_\lambda \as-\nonumber \\
&&\quad-\am \an \al \ast+\frac 12 \am \ant \al \ast+\frac 12 \amt \an \alt
\as)]
\label{18}
\eea
One sees that on the subgroup $SO(N)$ $(\lambda^a=-\lambda^{a,T})$  this
anomaly vanishes. The Wess--Zumino consistensy condition is obviously satisfied
because our anomaly differs from the standard one by the trivial 1-cocycle.

The additional $SO(N)$ invariance of the Wess--Zumino action $\aor(A,g)$
means that it depends in fact not on all the elements   of  $SU(N)$ but only
on the elements of the homogeneous space $SU(N)/SO(N)$. One can introduce
coordinates on this homogeneous space and express the Wess--Zumino action
in terms of these coordinates.

The natural coordinates are symmetric  and unitary matrices
\be
s=gg^T
\label{19}
\ee
This choice is suggested by the form of the mass term in the regularized
Lagrangian (\ref{14}). As follows from eq.(\ref{5}) after the gauge
transformation it  depends only on the combination  $gg^T$. The gauge group
transforms the coordinates $s$ in the following manner
\be
s\rightarrow  g^{-1}sg^{-1,T}.
\ee
 In terms of these coordinates the Wess--Zumino action looks
as follows:
\bea
\aor&=&\int d^4x\,[\frac 12d^{-1}\kappa (s)-\frac {i}{48\pi^2}
\epsilon^{\mu \nu \lambda \sigma }\tr[(\partial_\mu \an \al
+ \am \partial_\nu \al +\am \an \al-\nonumber \\
& &-\frac 12 \partial_\mu
\an s\alt s^{-1}- \frac 12 s \amt s^{-1} \partial_\nu \al- \am s\ant
s^{-1}\al)s_\sigma-\nonumber \\
& &-\frac 12 \am s_\nu \al s_\sigma +\frac 12 (s\amt s^{-1}\an-\am s\ant
s^{-1})s_\lambda s_\sigma - \am s_\nu s_\lambda s_\sigma \nonumber \\
& &+\partial_\mu \an \al s\ast s^{-1}+ \am \partial_\nu \al s\ast s^{-1} +
\am \an \al s\ast s^{-1}-\nonumber \\
& &-\frac 14 \am s \ant s^{-1}\al s\ast s^{-1}-\alpha_0 (A)]]
\label{20}
\eea
where $s_\mu=\partial_\mu ss^{-1}$.

The derivation is straightforward but some comments are  in order. Using the
equality
\be
g_\mu^T=s^{-1}(s_\mu-g_\mu)s
\label{21}
\ee
we express $g_\mu^T$ in terms of $g_\mu$ and $s_\mu$ and then comparing  the
terms of a given order in $\am$ and applying again eq.(\ref{21}) we find the
expression (\ref{20}). This action may be used for the construction  of the
symplectic form defining the integration messure in   the path integral.
It is worthwhile to emphasize that contrary to  the standard
case the action (\ref{20}) depends not only on the chiral current
$\partial_\mu ss^{-1}$, belonging  to the Lie algebra of the group, but also
on the coordinates of the homogeneous space $SU(N)/SO(N)$. It  may be of
importance for analyzing possible stationary points of the effective action.

\section {Anomalous constraints commutator}

In this section  we shall calculate the 2-cocycle
associated to the Wess--Zumino action (\ref{11}). This 2-cocycle appears  as
the Schwinger term  in the constraints commutator and can be calculated either
by direct summation of the Feynman diagrams \cite{KSS,Jo} or by using the
path integral representation for the commutator \cite{AMF}. We use the second
approach.  According to the Bjorken--Johnson--Low (BJL) formula the matrix
element of the equal time commutator may be expressed in terms of the
expectation value of $T$-product as  follows:
\be
\lim_{q_0 \rightarrow\infty} q_0 \int dt'\,\hbox{e}^{iq_0(t'-t) }\langle
\tilde{\varphi}
|\,TA(x,t')B(y,t)\,|\varphi  \rangle =i\langle \tilde{\varphi}
|\,[A(x,t),B(y,t)]\,|\varphi  \rangle
\label{22}
\ee

For the expectation value of $T$--product one can write the representation
in terms of the path integral
\be
\langle \tilde{\varphi}
|\,TA(x,t')B(y,t)\,|\varphi  \rangle =\int d\mu\,\hbox{e}^{iS}A(x,t')B(y,t)
\label{23}
\ee
Here it is understood that the integration goes over the fields satisfying the
boundary conditions corresponding to the initial and final states $|\varphi
\rangle$ and $\langle \tilde{\varphi} |$. Following the approach of
\cite{AMF} we can consider the chiral $SU(N)$ Yang--Mills model in the
Hamiltonian gauge $A_0=0$. In this gauge the $S$--matrix element can be
written as the path integral
\be
\langle \alpha |\beta \rangle=\int d\mu \,\delta  (A_0) \hbox{e}^{iS},
\label{24}
\ee
where in the first order formalism
\bea
S&=&\int d^4x\,[E_i^a\dot{A}_i^a-\frac 12 {(E_i^a})^2-\frac 14{(F_{ij}^a)}^2+
A_0^a G^a+\nonumber \\
&&+i\bar{\psi} \gamma_0 \partial_0\psi-i\bar{\psi} \gamma_i
(\partial_i-A_i)\psi ]
\label{26}
\eea

In the nonanomalous  case the constraints $G^a$ form a Lie algebra
\be
[G^a({\bf x}),G^b({\bf y})]=if^{abc}G^c({\bf y})\delta ({\bf x}-{\bf y})
\label{27}
\ee

However as  was shown in refs. \cite{AMF,KSS,Jo} in the anomalous theory this
relation is violated and the Schwinger term arises.

To calculate this Schwinger term we make the gauge transformation of the
variables in the integral (\ref{24}). The transformed integral   may be
written in the form
\be
\langle \tilde{\varphi} |\varphi \rangle=\int d\mu \, \delta  (A_0)
\hbox{e}^{iS}\exp \{-i\int  d^4x\,g_0^a G^a(x)+i\aor
(A,g)\bigm|_{A_0=-g_0}\}
\label{28}
\ee
Here the 1-cocycle arises due to the noninvariance of the regularization in
accordance  with eq.(\ref{6}).

Using the  representation for  the chiral field $g$: $g=\hbox{e}^u$ and taking
into  account  that the integral (\ref{28}) does not  depend on $g$ we can
put equal to zero variation of this integral over  $u$. To the second order
in $u$ one has:
\bea
&&\frac 12 \int d^4x d^4y \,\langle \tilde{\varphi} |T\widetilde{G}^a(x)
\widetilde{G}^b(y)|\varphi  \rangle \partial_0 u^a(x) \partial_0
u^b(y)+\nonumber \\
&&+\frac i2 \int d^4x \,f^{abc}u^a(x) \partial_0 u^b(y)\langle \tilde{\varphi}
|T\widetilde{G}^a(x) |\varphi  \rangle+\nonumber \\
&&+\frac {1}{48\pi^2} \int d^4x\,\langle\tilde{\varphi}| \tr(\epsilon^{ijk}
\partial_i A_j \{\partial_k u(x),  \partial_0 u(x)\}) |\varphi  \rangle =0 .
\label{29}
\eea
Here we introduced the notation
\bea
\widetilde{G}^a(x)&=&G^a(x)-\frac {i}{48\pi^2}\epsilon_{ijk}\tr[(\lambda^a  +
\lambda^{a,T})(A_i \partial_j A_k+\partial_i A_jA_k+A_iA_jA_k-A_i A_j^T
A_k)\nonumber   \\
& &\quad-\lambda^a \{\partial_i A_j,A_k^T\}]
\label{30}
\eea
In the process of derivation  of eqs.(\ref{29}), (\ref{30}) we used the
explicit
form of $\aor$ (\ref{11}), (\ref{16}) and made the shift of the variables
$E_i^a$ \be E_i^a  \rightarrow E_i^a+\frac {i}{48\pi^2}\epsilon_{ijk}\tr
\lambda^a(\{A_j,g_k\}+g\{A_j^g,A^{T,g}_k\}g^{-1}-\{A_j,A_k^T\})
\label{31}
\ee
In eq.(\ref{29}) we kept only the terms nonvanishing in the BJL limit.

To get the expression for the commutator of  $\widetilde{G}$ we apply to
eq.(\ref{29}) the operator:
\be
\lim_{(p_0-q_0)\rightarrow\infty}\frac {p_0-q_0}{p_0q_0}\int
dx_0dy_0\,\hbox{e}^{ip_0x_0+iq_0y_0} \frac {\delta}{\delta  u^a(x)}\frac
{\delta}{\delta  u^b(y)}
\label{32}
\ee
Taking the limit we  get  the result
\be
[\widetilde{G}^a({\bf x}),\widetilde{G}^b({\bf
y})]=if^{abc}\widetilde{G}^c({\bf y})\delta ({\bf x}-{\bf y})
-\frac {1}{24\pi^2} \epsilon_{ijk}\tr (\partial_i A_j\{\lambda^a
,\lambda^b\})\partial_k^x \delta ({\bf x}-{\bf y}).
\label{33}
\ee

Let us note that the commutator  of $\widetilde{G}$ (\ref{33}) coincides
with the analogous commutator obtained  in ref.\cite{AMF} with the different
Wess--Zumino  action. However the definition  of  $\widetilde{G}$ in our case
is different. If one comes  back to the constraints $G$ one gets
\be
[G^a ({\bf x}),G^b ({\bf y})]=if^{abc}G^c({\bf y})\delta
({\bf x}-{\bf y})+a_{2,ort}^{ab} (A;{\bf x},{\bf y})
\label{34}
\ee
Here $a_{2,ort}^{ab} $  is the ultralocal 2-cocycle
\bea
&&a_{2,ort}^{ab} (A;{\bf x},{\bf y})=-\frac {1}{48\pi^2}\epsilon_{ijk}\tr
([\lambda^a+\lambda^{a,T},\lambda^b+\lambda^{b,T}]\times \nonumber\\
&&\quad\times(A_i\partial_jA_k+
\partial_iA_jA_k
+A_iA_jA_k-A_iA_j^TA_k+A_i^T\partial_jA_k+\partial_iA_j^TA_k)+\nonumber \\
&&\quad+(\lambda^a+\lambda^{a,T})(\partial_iA_j-\partial_iA_j^T-A_iA_j^T-A_i^T
A_j) (\lambda^b+\lambda^{b,T})A_k-\nonumber \\
&&\quad-(\lambda^b+\lambda^{b,T})(\partial_iA_j-\partial_iA_j^T-A_iA_j^T-A_i^TA_j)
(\lambda^a+\lambda^{a,T})A_k )
\label{35}
\eea

This cocycle obviosly differs from the one obtaned in
ref.\cite{AMF}--\cite{Jo}. In particular it vanishes if at least one of the
constraints $G^a$ corresponds
to the subgroup $SO(N)$. We note that the adding to 1-cocycle any trivial
1-cocycle having topological nature does not change the commutator of
modified constraints  $\widetilde{G}$.

\section {Conclusion}
In this paper we showed that the anomalous $SU(N)$ gauge theory may be
quantized in such a  way that the resulting effective action possesses the
residual $SO(N)$ symmetry. This action is described in a natural  way in
terms of the coordinates of the homogeneous space $SU(N)/SO(N)$. An
analogous construction may be  carried out for other gauge groups and
representations having nonanomalous subgroups. A trivial local 1-cocycle  in
this case can be calculated using the same equation $\alpha (A,h)=0$ where
$h$ is an element of the nonanomalous subgroup. The next problem is to try to
investigate the physical content of this theory.

If one  follows the approach of Faddeev and Shatashvili \cite{FS86}
to quantization of anomalous theories one should add to the original classical
action the  Wess--Zumino term. In general the physical content of  the model
depends on the particular form of modified Lagrangian.

The effective action we got and the constraints  algebra differ from the ones
obtained by Faddeev and Shatashvili. It would be very interesting  to
investigate if the number of physical degrees of freedom is the same or
different in both cases. One possibility is that the variables corresponding to
the $SO(N)$ subgroup in the Faddeev--Shatashvili action are not dynamical and
integrating them  out one would get our action. If it is not  the case that
means these two approaches  lead to physically different  models.

To analize the physical content of the theory  one needs to develop the
expansion of the path integral near some stationary point. At present it is
an open problem, because in the four-dimensional case such a solution is not
known, and in two dimensional models this effect is absent.

Let us note  that in our case the effective action  depends not only on the
chiral currents and in principle allows constant solution for
coordinates of the homogeneous space $SU(N)/SO(N)$.

Acknowledgements: One of the authors (S.F.) would like to thank Professor
P.K.Mitter and Laboratoire de
Physique Th\'eorique et des Hautes \'Energies for hospitality.

\end{document}